\documentstyle[12pt,a4wide]{article}
\newcommand{\be}{\begin{equation}}
\newcommand{\ee}{\end{equation}}
\newcommand{\bea}{\begin{eqnarray}}
\newcommand{\eea}{\end{eqnarray}}
\newcommand{\bean}{\begin{eqnarray*}}

\newcommand{\eean}{\end{eqnarray*}}
\font\upright=cmu10 scaled\magstep1 \font\sans=cmss10
\newcommand{\ssf}{\sans}
\newcommand{\stroke}{\vrule height8pt width0.4pt depth-0.1pt}
\newcommand{\Z}{\hbox{\upright\rlap{\ssf Z}\kern 2.7pt {\ssf Z}}}

\newcommand{\C}{{\rlap{\rlap{C}\kern 3.8pt\stroke}\phantom{C}}}
\newcommand{\R}{\hbox{\upright\rlap{I}\kern 1.7pt R}}
\newcommand{\CP}{\C{\upright\rlap{I}\kern 1.5pt P}}
\newcommand{\PP}{\hbox{\upright\rlap{I}\kern 1.5pt P}}

\newcommand{\identity}{{\upright\rlap{1}\kern 2.0pt 1}}
\newcommand{\identito}{{1\hspace{-0.11cm}1}}

\newcommand{\HH}{\mbox{\hbox{\upright\rlap{I}\kern 1.7pt H}}}

\newcommand{\xib}{{\bar \xi}}

\newcommand{\fr}{\frac}

\newcommand{\ra}{\rightarrow}
\newcommand{\al}{\alpha}
\newcommand{\bt}{\beta}
\newcommand{\pr}{\partial}

\newcommand{\dg}{\dagger}

\newcommand{\vphi}{{\varphi}}

\input{epsf}
\begin{document}

\title{\bf Platonic Gravitating Skyrmions}
\vspace{1.5truecm}
\author{
{\bf Theodora Ioannidou$^a${\footnote{{\it Email}: ti3@auth.gr}},
 \bf Burkhard Kleihaus$^b${\footnote{{\it Email}: kleihaus@theorie.physik.uni-oldenburg.de}}
and Jutta Kunz$^b${\footnote{{\it Email}: kunz@theorie.physik.uni-oldenburg.de}}
}\\
$^a$ Mathematics Division, School of Technology\\
Aristotle University of Thessaloniki \\
Thessaloniki 54124,  Greece\\
$^b$ Institut f\"ur  Physik, Universit\"at Oldenburg, Postfach 2503\\
D-26111 Oldenburg, Germany
}

\vspace{1.5truecm}

\date{\today}

\maketitle

\vspace{1.0truecm}

\begin{abstract}
We construct globally regular gravitating Skyrmions,
which possess only discrete symmetries.
In particular, we present tetrahedral and cubic Skyrmions.
The $SU(2)$ Skyrme field is parametrized by an improved harmonic map
ansatz. Consistency then requires also a restricted ansatz for the metric.
The numerical solutions obtained within this approximation
are compared to those obtained in dilaton gravity.
\end{abstract}


\section{Introduction}

Nonlinear field theories coupled to gravity 
lead to globally regular gravitating configurations \cite{review}.
Moreover, also black hole solutions with
nonlinear hair arise \cite{review}.
These black hole solutions are asymptotically flat and
possess a regular event horizon.
Outside their horizon they retain the features of the
corresponding gravitating solitons,
and may thus be viewed as bound states of solitons and 
Schwarzschild black holes \cite{iso}.

In the Einstein-Skyrme model the nonlinear chiral field theory
describing baryons and nuclei in terms of solitons (so-called Skyrmions)
is coupled to gravity.
Static spherically symmetric $SU(2)$ gravitating Skyrmions and black holes
with Skyrmion hair \cite{GKK}-\cite{BC} exhibit a characteristic
dependence on the coupling parameter:
two branches of solutions merge and end 
at a maximal value of the coupling parameter,
and only the solutions on the lower branch are
classically stable.
Spherically symmetric $SU(N)$ gravitating Skyrmions and
black holes have been analyzed similarly
\cite{KKS,BHIZ}.

Recently, static axially symmetric $SU(2)$ gravitating Skyrmions 
and black holes with Skyrmion hair
have been constructed numerically \cite{Japanese},
while approximations to axially symmetric gravitating $SU(3)$ Skyrmions
have been obtained in \cite{IKZ1}.
The hairy black holes represent further examples demonstrating
that Israel's theorem does not generalize to theories with non-Abelian fields
\cite{KKbh}.

On the other hand, in flat space also Skyrmions 
with no rotational but only discrete symmetries have been constructed
\cite{BTC}.
Among them are solutions with the symmetries of the platonic solids,
to which we refer as platonic Skyrmions.
Besides the exact numerical solutions, also approximate solutions
have been obtained.
Such approximate Skyrmion solutions with baryon number $B$
have been constructed using rational maps of degree $B$ between Riemann 
spheres \cite{HMS}, 
as well as an improved harmonic map ansatz \cite{IKZ2}.

In this paper we consider gravitating Skyrmion configurations
with only platonic symmetries. 
In particular, we focus on
configurations with tetrahedral and cubic symmetry,
possessing baryon number $B=3$ and $B=4$, respectively.

Recall that the $SU(2)$ Einstein-Skyrme action reads
 \be
  S=\int \left[
  \fr{R}{16\pi G}
 +\fr{\kappa^2}{4} \mbox{Tr}\left(K_{\mu}\,K^\mu \right) 
 +\fr{1}{32e^2}\mbox{Tr}\left(
     \left[K_\mu,K_\nu\right]\left[K^\mu,K^\nu\right]
                        \right)
	\right]
                \sqrt{-g}\, d^4x \ , 
  \label{ac}
  \ee
where $R$ is the curvature scalar,
$g$ denotes the determinant of the metric, 
the $SU(2)$ Skyrme field $U$ enters via $K_\mu=\pr_\mu U U^{-1}$,
$G$ represents Newton's constant,
and $\kappa$ and $e$ are the Skyrme model coupling constants.

In order for finite-energy configurations to exist the Skyrme field
must tend to a constant matrix at spatial infinity:  
$U \ra \identity$ as
$|x^\mu| \ra \infty$.
This effectively compactifies the three-dimensional Euclidean space
into $S^3$ and implies that the Skyrme fields 
can be considered as maps from $S^3$ into $SU(2)$.

Variation of action (\ref{ac}) with respect to the metric 
$g^{\mu \nu}$ leads to the Einstein equations
\be
G_{\mu \nu}=R_{\mu \nu}-\fr{1}{2}g_{\mu \nu}R=8\pi G T_{\mu \nu} \ , 
\label{E}
\ee
with stress-energy tensor 
\be
T_{\mu\nu}\!\!=
-\fr{\kappa^2}{2}
\mbox{Tr}\!\left(K_\mu K_\nu-\fr{1}{2}g_{\mu \nu} K_\al K^\al \right)
- \fr{1}{8e^2} 
\mbox{Tr}\!\left(
g^{\al \bt}\left[K_\mu,K_\al\right]\left[K_\nu,K_\bt\right]-
\fr{1}{4}g_{\mu \nu} \left[K_\al,K_\bt\right]\left[K^\al,K^\bt\right]
\right).
\label{T}
\ee
Variation with respect to the Skyrme field
leads to the field equation
\be
\nabla_\mu \left(\kappa^2 K^\mu 
 + \frac{1}{4e^2}[K_\nu, [K^\mu,K^\nu]] \right) = 0 \ .
\label{Ueq}
\ee

The Einstein-Skyrme system has a topological current
\begin{equation}
B^\mu = \frac{1}{\sqrt{-g}} \frac{1}{24\pi^2}
\epsilon^{\mu \nu \alpha \beta} {\rm Tr}\, \left( K_\nu K_\alpha K_\beta \right)
\ , \end{equation}
which corresponds to the baryon current,
yielding the baryon number $B$ 
\begin{equation}
B = \int \sqrt{-g} B^0 d^3 x \ .
\label{B}
\end{equation}

For comparison, we also consider Skyrme-dilaton theory
with action 
\be
S = \int\left[ -\fr{1}{2} \partial_\mu \phi \partial^\mu \phi
 +\fr{\kappa^2}{4} \mbox{Tr}\left(K_{\mu}\,K^\mu \right)
 +\fr{1}{32e^2}e^{-2\gamma \phi} \mbox{Tr}\left(
     \left[K_\mu,K_\nu\right]\left[K^\mu,K^\nu\right]
                        \right)
        \right]
                \, d^4x  \ ,
  \label{ac-dil}
\ee
with $\gamma = \sqrt{4\pi G}$,
and indices are lowered and raised by the flat Minkowski metric.

The idea of the harmonic map ansatz for $SU(N)$ Skyrmions \cite{IPZ1}
(which is the generalisation of the rational map ansatz of 
Houghton et al.~\cite{HMS}) 
involves the separation of the radial and angular 
dependence of the fields as
\be
U=e^{2ih(r)\left(P-\identito/N \right)}
=e^{-2ih(r)/N}\left[\identity+\left(e^{2ih(r)}-1\right)P\right]
\ . \label{U}
\ee
Here $h(r)$ is the corresponding profile function
and $P$ is a $N\times N$ hermitian projector, which depends only on the
angular variables $(\xi,\xib)$, 
where $\xi$ is the Riemann sphere variable given by 
$\xi=e^{i\vphi} \tan(\theta/2)$ in terms of the usual spherical
coordinates $r,\theta,\vphi$.
Note that the matrix $P$ can be thought of as a mapping from $S^2$ into 
$CP^{N-1}$. Thus, $P$ can be written as
\be
P(V)=\fr{V \otimes V^\dg}{|V|^2} \ ,
\label{for}
\ee
where $V$ is a $N$-component complex vector (depending on $\xi$). 

The improved harmonic map ansatz is obtained by allowing the profile
function $h$ to depend on all spatial coordinates $r,\xi,\xib$.
As has been shown in \cite{IKZ2}, this ansatz leads to a better approximation
of the Skyrmion energy in flat space.

For $SU(2)$ Skyrmions the harmonic map ansatz can be related to 
rational maps $R(\xi)$ by
\be
V=\left(R(\xi),1\right)^t \ , 
\ee
where $R(\xi)=p(\xi)/q(\xi)$, and $p(\xi), \,q(\xi)$ are polynomials 
of $\xi$.
Defining  the unit vector $\hat{\vec{n}}_R$ by 
\be
\hat{\vec{n}}_R = \fr{1}{1+|R|^2}
\left( R+\bar{R}\ , \ -i(R-\bar{R}) \ , \ 1-|R|^2 \right) \ , 
\ee
the Skyrme field $U$ takes the simple form
\be
U = \cos(h) \identity +i \sin(h) \hat{\vec{n}}_R \cdot \vec{\tau} \ , 
\label{UR}
\ee
with the vector of Pauli matrices $\vec{\tau}=(\tau_1\ ,\ \tau_2\ ,\ \tau_3)$.

\section{Metric Ansatz}

To obtain static solutions without rotational symmetries 
let us first consider the following ansatz for the metric 
\begin{equation}
ds^2=-fdt^2+\fr{1}{f}
\left( m_1 dr^2+ m_2 r^2 d\theta^2 +l r^2 \sin^2\theta d\vphi^2\right)  \ , 
 \label{s}
\end{equation}
where $f$, $m_1$, $m_2$ and $l$ are functions of $r$, $\theta$ and $\vphi$.
In the case of axial symmetry, $m_1=m_2=m$ \cite{KK}.

Insertion of the Skyrme field (\ref{UR}) and the metric (\ref{s}) 
in the action (\ref{ac}),
and subsequent variation of the action with respect to the 
Skyrme profile function $h$ 
yields a second order partial differential equation 
(PDE) for $h$.
Similarly, PDEs for the metric functions are obtained,
which are equivalent to those obtained from the general Einstein
equations (\ref{E}) after insertion of the Skyrme field 
and the metric.

It now turns out, that this coupled system of PDEs
does not possess a solution in general.
While surprising at first, this fact has the following reason:
The Einstein tensor $G_{\mu\nu}$ is defined so that its covariant 
divergence vanishes, 
\be
\nabla_{\mu} G^{\mu\nu} = 0\ . 
\ee
Consequently, the Einstein equations require
\be
\nabla_{\mu} T^{\mu\nu} = 0\ ,
\label{Ein}
\ee
provided there exists a solution.
Insertion of the Skyrme field (\ref{UR})
and the metric (\ref{s}) yields, however,
\bea
\nabla_{\mu} T^{\mu 0} = 0\ ,\ \ \ && \nabla_{\mu} T^{\mu r} = 0 \ ,
\nonumber\\ 
\nabla_{\mu} T^{\mu \theta} \neq 0\ , \ \ \ &&
\nabla_{\mu} T^{\mu \vphi} \neq 0\ .
\eea
Consequently, not all Einstein equations can be satisfied.
In the axially symmetric case, when $m_1=m_2$ and 
the Skyrme and metric functions do not depend on $\vphi$,
we can achieve $\nabla_{\mu} T^{\mu \vphi}=0$, 
but still $\nabla_{\mu} T^{\mu \theta} \neq 0$.
Only in the spherically symmetric case,
when $m_1=m_2=l$ and
the Skyrme and metric functions depend only on $r$,
all equations in (\ref{Ein}) are satisfied.
In this case, however, 
the ansatz for the Skyrmion field leads to an exact solution.

This problem can be traced back to the fact that the harmonic map ansatz
for the Skyrme field is too restrictive. 
Indeed, for a general ansatz for the Skyrme field $U$
involving three functions 
$h_i(r,\theta,\vphi)$
\be
U= \cos(h_1) \identity + i \sin(h_1)
\left[ \sin(h_2)(\cos(h_3) \tau_1 + \sin(h_3) \tau_2)
       + \cos(h_2) \tau_3 \right]
\label{genans} \ee
all equations in (\ref{Ein}) are satisfied.

While aiming at the numerical construction 
of exact platonic gravitating Skyrmions,
in view of the complexity of the coupled
Einstein-Skyrme equations
we first want to obtain simpler approximate solutions,
based on the improved harmonic map ansatz.

We therefore argue as follows: since we employ only an approximate
ansatz for the Skyrme field, we should restrict also to an approximate 
ansatz for the metric, compatible with the ansatz for the Skyrme field.
An appropriate ansatz for the metric is given by
\be
ds^2 = -f dt^2 
 +\frac{l}{f}\left( dr^2 + r^2 d\theta^2 +r^2 \sin^2\theta d\vphi^2\right)
\ , \label{ma2} \ee
where we allow only for two metric functions $f$ and $l$
(as in the spherically symmetric case), 
which, however, depend on all three coordinates,
like the Skyrme function $h$ of the improved harmonic map ansatz.

We then derive the set of three coupled PDEs 
as variational equations from the Einstein-Skyrme action (\ref{ac}),
after the second order derivatives of the metric functions have 
been eliminated by intergration by parts. 
We refer to this approximation as ``$f-l$-approximation''.

We note that a further restriction of the metric obtained
by setting $l=1$, leads to solutions of the 
Skyrme-dilaton model with action (\ref{ac-dil}). 
In this case the dilaton can be expressed by the metric function $f$,
\be
\phi = -\fr{1}{2 \gamma}\log(f) \ , \ \ \
\gamma = \sqrt{4 \pi G}
\ee 
We refer to this approximation as ``dilaton-approximation''.

To obtain an estimate for the quality of the approximation, 
we can substitute the approximate solutions
in the full set of Einstein and matter equations, 
to see how strongly these are violated.

\section{Numerical Solutions}

\subsection{Parameters and Boundary Conditions}

Introducing the dimensionless radial coordinate $x=\kappa e r$ and 
coupling parameter $\alpha = 4 \pi G\kappa^2$, the action (\ref{ac}) becomes
\be
  S=\fr{\kappa}{e^2}
  \int \left[
  \fr{R}{4\alpha}
 +\fr{1}{4} \mbox{Tr}\left(K_{\mu}\,K^\mu \right) 
 +\fr{1}{32}\mbox{Tr}\left(
     \left[K_\mu,K_\nu\right]\left[K^\mu,K^\nu\right]
                        \right)
	\right]
                \sqrt{-g}\, d^4x \ ,
  \label{ac_n}
\ee
and the Einstein equations read $G_{\mu\nu} = 2 \alpha T_{\mu\nu}$.
Thus the solutions depend only on the coupling parameter $\alpha$ and
the chosen rational map.
Likewise, in Skyrme-dilaton theory, the solutions depend only on
the dimensionless coupling parameter $\alpha = \gamma^2 \kappa^2$
and the chosen rational map.


For Skyrmions with axial symmetry rational maps of degree $B$ 
are simply given by $R_B = \xi^B$, while
for Skyrmions with platonic symmetries these maps are more complicated.
In particular, for tetrahedral $B=3$ Skyrmions
and cubic $B=4$ Skyrmions these maps are given by
\be
R_{\rm tetra} = \fr{\sqrt{3} \xi^2 +1}{\xi(\xi^2-\sqrt{3})} 
\ee
and 
\be
R_{\rm cube} = \fr{\xi^4+2\sqrt{3}i\xi^2+1}{\xi^4-2\sqrt{3}i\xi^2+1} \ ,
\ee
respectively.


In order to map the infinite range of the radial variable $x$ to the finite 
interval $[0,1]$ we introduce the compactified radial variable 
$\bar{x} = x/(1+x)$.

For axially symmetric solutions the Skyrme and metric functions 
depend only on the coordinates $\bar{x}$ and $\theta$. 
Due to the reflection symmetry, $ z \leftrightarrow -z$, it is sufficient
to construct solutions for 
$0\leq \bar{x} \leq 1$, $0\leq \theta \leq \pi/2$.
The boundary conditions at the origin are 
\be
h(0) = \pi \ , \ \ \partial_x f|_0 = 0 \ , \ \ \partial_x l|_0 = 0  \ . 
\ee
Asymptotically the Skyrme field tends to the unit matrix and 
the metric approaches the Minkowski metric, i.e.
\be
h(\infty) \to 0 \ , \ \ f(\infty) \to  1  \ , \ \  l(\infty)\to 1 \ .
\ee
On the $z$-axis ($\theta=0$) and in the $xy$-plane ($\theta=\pi/2$)
the boundary conditions follow from regularity and reflection symmetry,
respectively,
\be 
\partial_\theta h = 0 \ , \ \ 
\partial_\theta f = 0 \ , \ \ 
\partial_\theta l = 0 \ . 
\ee

For solutions with discrete symmetries the Skyrme and metric functions
depend on all three coordinates $x,\theta,\vphi$.
The tetrahedral symmetry of the $B=3$ solution allows
to restrict to the domain of integration to
$0\leq \bar{x} \leq 1$, $0\leq \theta \leq \pi$, $0\leq \vphi \leq \pi/2$.
Similarly, the cubic symmetry of the $B=4$ solution allows
to restrict to the domain of integration to
$0\leq \bar{x} \leq 1$, $0\leq \theta \leq \pi/2$, $0\leq \vphi \leq \pi/2$.
The boundary conditions at the origin, at infinity, on the $z$-axis and 
in the $xy$-plane are the same as for the axially symmetric solutions.
The remaining boundary conditions at $\vphi=0$ and $\vphi=\pi/2$
follow from the platonic symmetries, i.e.
\be 
\partial_\vphi h = 0 \ , \ \ 
\partial_\vphi f = 0 \ , \ \ 
\partial_\vphi l = 0 \ . 
\ee

\subsection{Numerical Results}

Solutions are constructed with help of the
software package FIDISOL \cite{fidisol} based on the Newton-Raphson
algorithm.
Typical grids contain $70\times 30$ points for
the axially symmetric solutions and $70 \times 25 \times 25$ points
for the platonic solutions.
The estimated relative errors are approximately $\approx 0.1$\%,
except close to $\alpha_{\rm max}$, where they become as large as 
$1$\%.

We have constructed gravitating Skyrmions with baryon number $B=2,3,4$
in the ``$f-l$-approximation'' and ``dilaton-approximation'',
and studied their dependence on the coupling parameter $\alpha$.
The $\alpha$-dependence of
the axially symmetric ($B=2,3,4$) and platonic ($B=3,4$) Skyrmions
is completely analogous to the $\alpha$-dependence of the
spherically symmetric $B=1$ Skyrmions \cite{DHS}-\cite{KKS}.

Gravitating Skyrmions 
exist only up to a maximal value of the coupling parameter,
$\alpha_{\rm max}$,
which depends on the specific rational map and on the approximation
(see Table 1).
\begin{table}
\begin{center}
\begin{tabular}{|c|c|c|}
\hline
\hline
& \multicolumn{2}{c|}{$\alpha_{\rm max}$}\\
\hline
rat. map & ``f-l" & ``dilaton" \\
\hline
$R_2$ & 0.0318    &  0.0294   \\
\hline
$R_3$       & 0.0266   &  0.0246     \\
$R_{tetra}$ & 0.0267   &  0.0248    \\
\hline
$R_4$      & 0.0231   & 0.0214     \\
$R_{cube}$ & 0.0234   & 0.0218 \\    
\hline
\hline
\end{tabular}
\caption{The maximal value of the coupling parameter, $\alpha_{\rm max}$,
for axial ($R_2$, $R_3$, $R_4$) and platonic ($R_{tetra}$, $R_{cube}$)
Skyrmions.}
\end{center}
\end{table}
When $\alpha$ is increased from zero a branch of gravitating
Skyrmions emerges from the corresponding flat space Skyrmion solution.
This first (lower) branch extends up to the maximal value $\alpha_{\rm max}$,
where it merges with a second (upper) branch of solutions. 
The upper branch then extends back to $\alpha=0$.

\begin{figure}
\centering
\epsfysize=5.5cm
\mbox{\epsffile{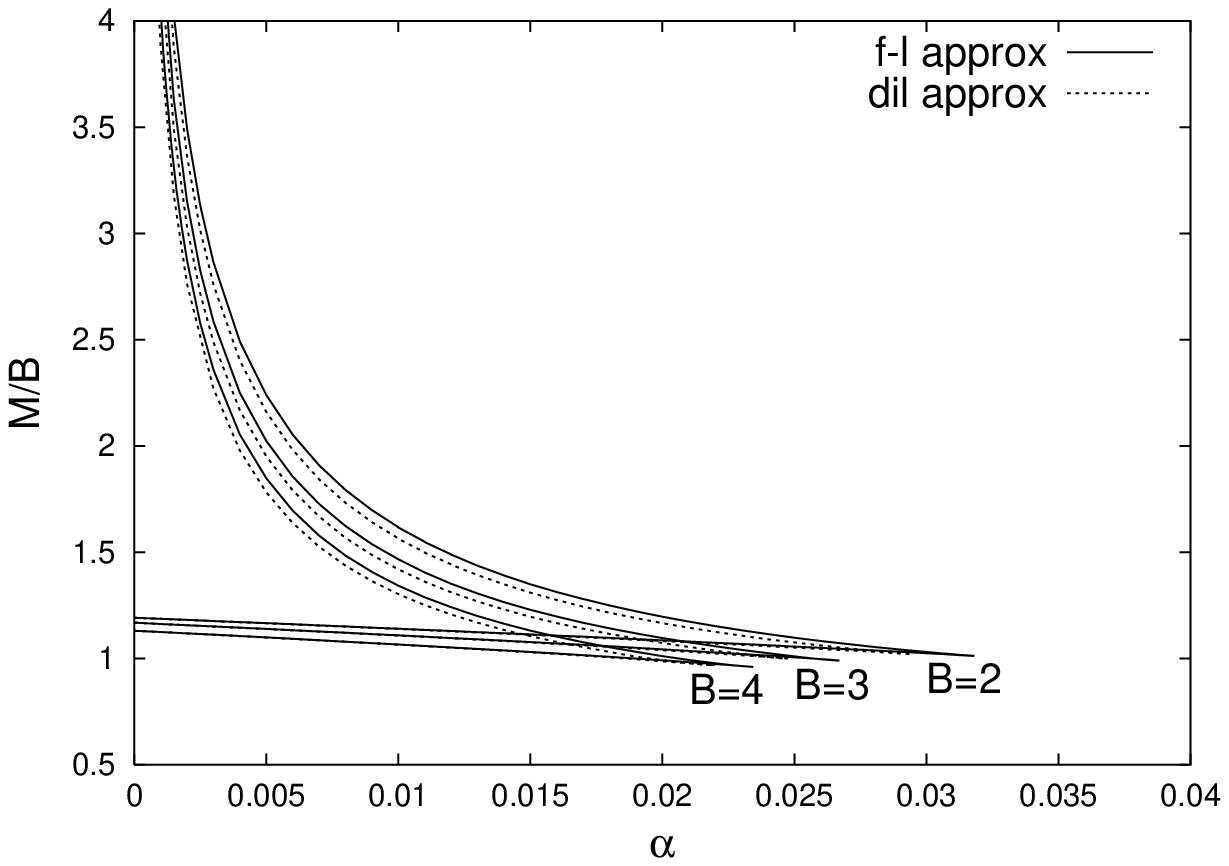}}
\epsfysize=5.5cm
\mbox{\epsffile{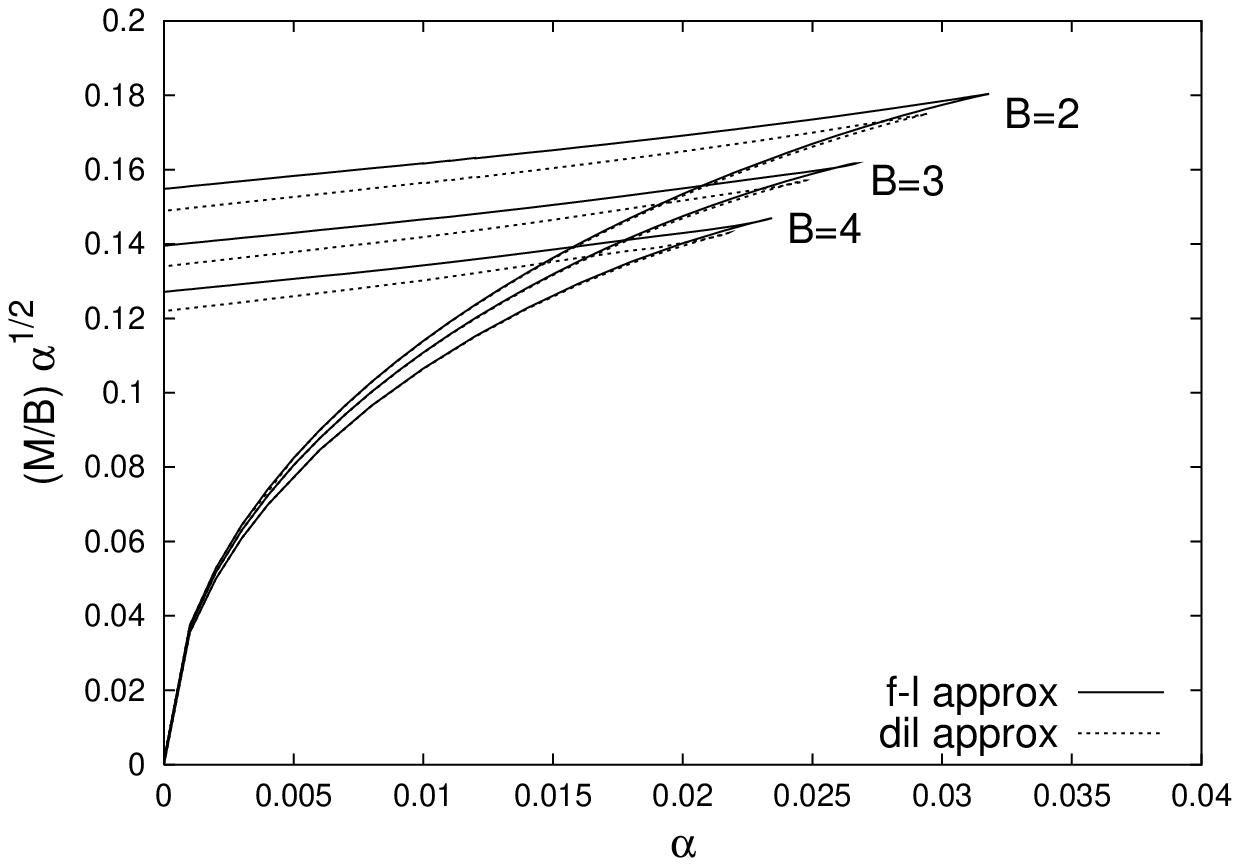}}
\caption{\label{Fig1} 
The dimensionless mass per baryon number $M/B$ (left) and the
scaled mass per baryon number $(M/B)\sqrt{\alpha}$ (right)
are shown as functions of the coupling parameter $\alpha$ 
for axial ($B=2$) and platonic ($B=3,4$) Skyrmions
in the ``$f-l$-approximation'' and the ``dilaton-approximation''.
}
\centering
\end{figure}

In Fig.~\ref{Fig1} we present the mass per baryon number 
as a function of $\alpha$ (left), 
for axially symmetric $B=2$ and platonic $B=3$ and $B=4$
Skyrmions.
On both branches the mass decreases with increasing $\alpha$.
But whereas the mass remains
finite in the limit $\alpha \to 0$ on the lower branch,
it diverges in this limit on the upper branch.
Thus on the upper branch the limit $\alpha \to 0$ 
does not correspond to a flat space limit, where gravity decouples.

To better understand the limit $\alpha \to 0$ on the
upper branch, we note that the coupling parameter $\alpha=4\pi G \kappa^2$
vanishes, either when $G$ vanishes while $\kappa$ remains constant,
or when $\kappa$ vanishes while $G$ remains constant.
The first case corresponds to the flat space limit of the
lower branch, while the second case corresponds
to the limit $\alpha \to 0$ of the upper branch.
Introducing the rescaled radial coordinate $\tilde{x} = x/\sqrt{\alpha}$ 
and the rescaled mass $\tilde{M} = M\sqrt{\alpha}$,
one observes, that the rescaled mass remains finite
in the limit $\alpha \to 0$ on the upper branch,
as illustrated in Fig.~\ref{Fig1} (right).

We note that the $\alpha$-dependence of the mass of the
gravitating Skyrmions solutions is almost the same 
in the ``$f-l$-approxima\-tion'' as in the ``dilaton-approximation''.
In the ``$f-l$-approximation'' the mass is slightly higher,
in particular, along the upper branch. Also,
$\alpha_{\rm max}$ is slightly larger in the ``$f-l$-approximation''.
The mass of the axially symmetric $B=3$ and $B=4$ Skyrmions is always
larger than the mass of the corresponding platonic Skyrmions.

\begin{figure}[b!]
\centering
\epsfysize=5.5cm
\mbox{\epsffile{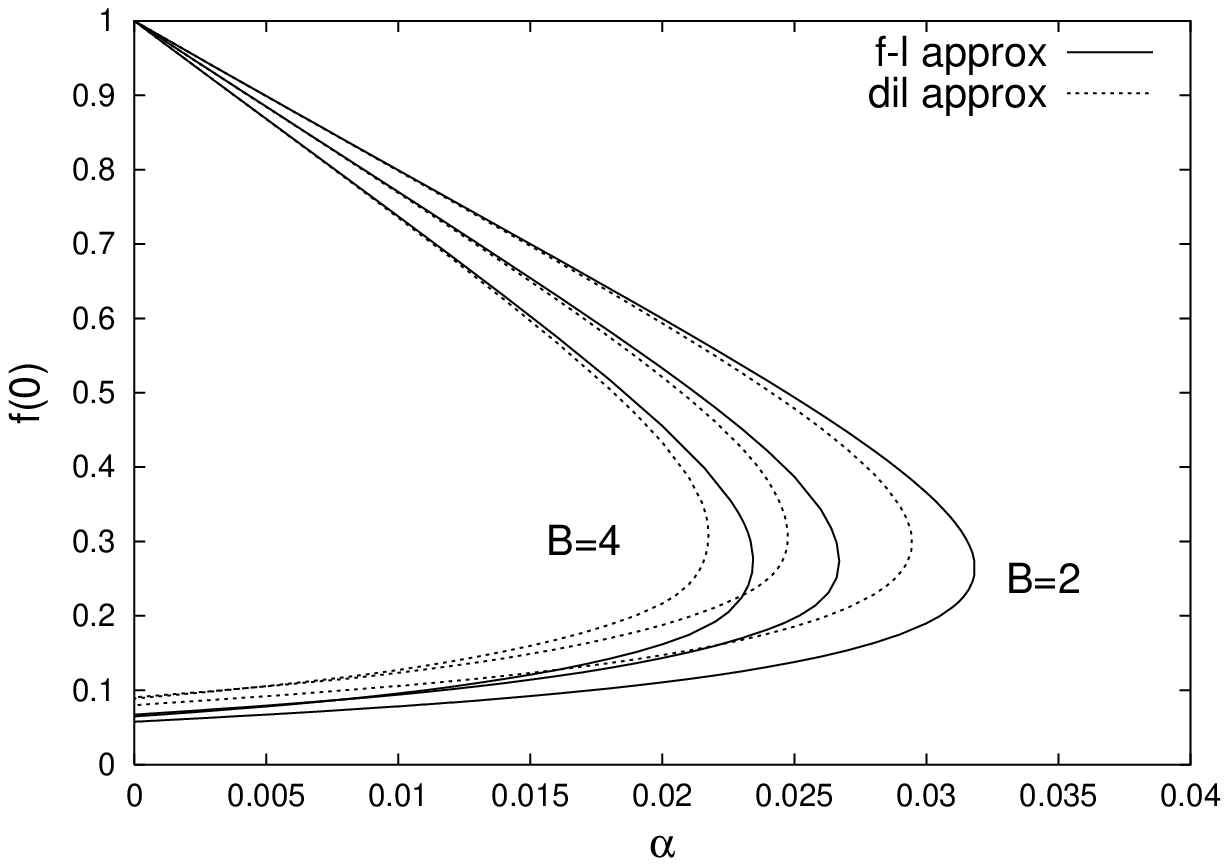}}
\epsfysize=5.5cm
\mbox{\epsffile{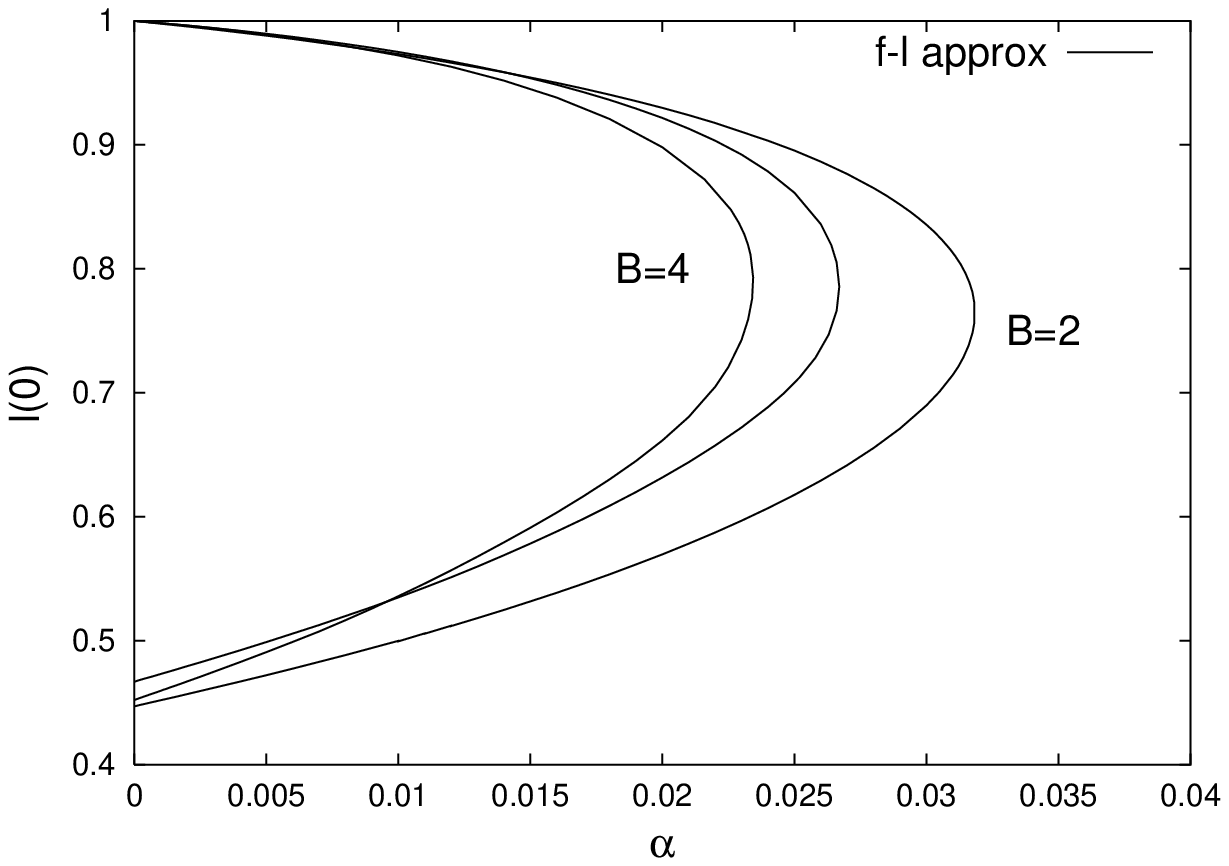}}
\caption{\label{Fig2} Same as \ref{Fig1} for the 
value of the functions $f$ and $l$ at the origin.
Note, that $l=1$ in the ``dilaton-approximation''.
}
\centering
\end{figure}

In Fig.~\ref{Fig2} we exhibit the value of the metric 
functions $f$ and $l$ at the origin 
for these axial ($B=2$) and platonic ($B=3,4$) Skyrmions.
We observe that $f(0)$ and $l(0)$ take finite values
in the limit $\alpha \to 0$ on the upper branch.

\begin{figure}[t!]
\parbox{\textwidth}{
\centerline{
\mbox{\epsfysize=9.0cm \epsffile{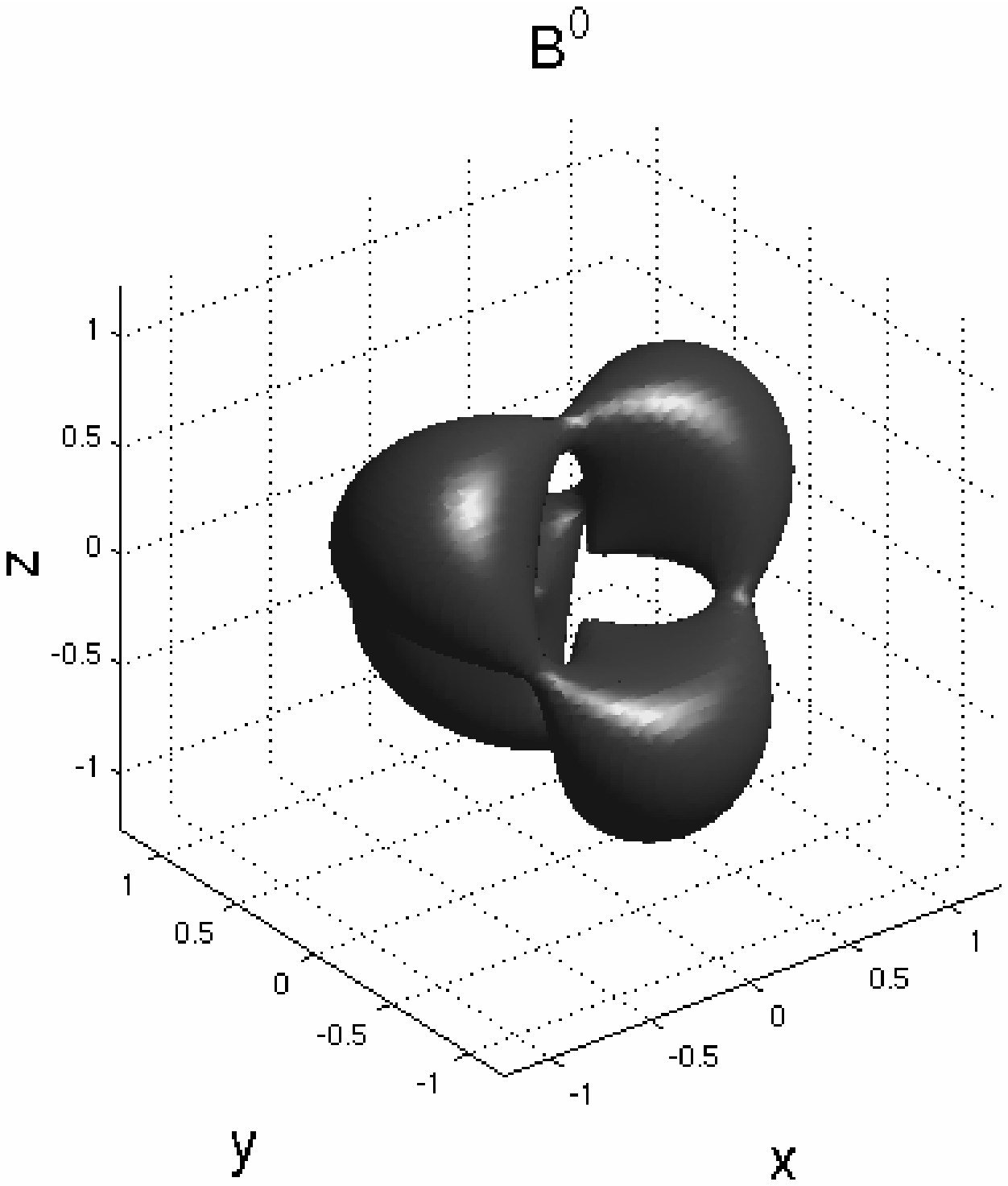} } \hspace{1cm}
\mbox{\epsfysize=9.0cm \epsffile{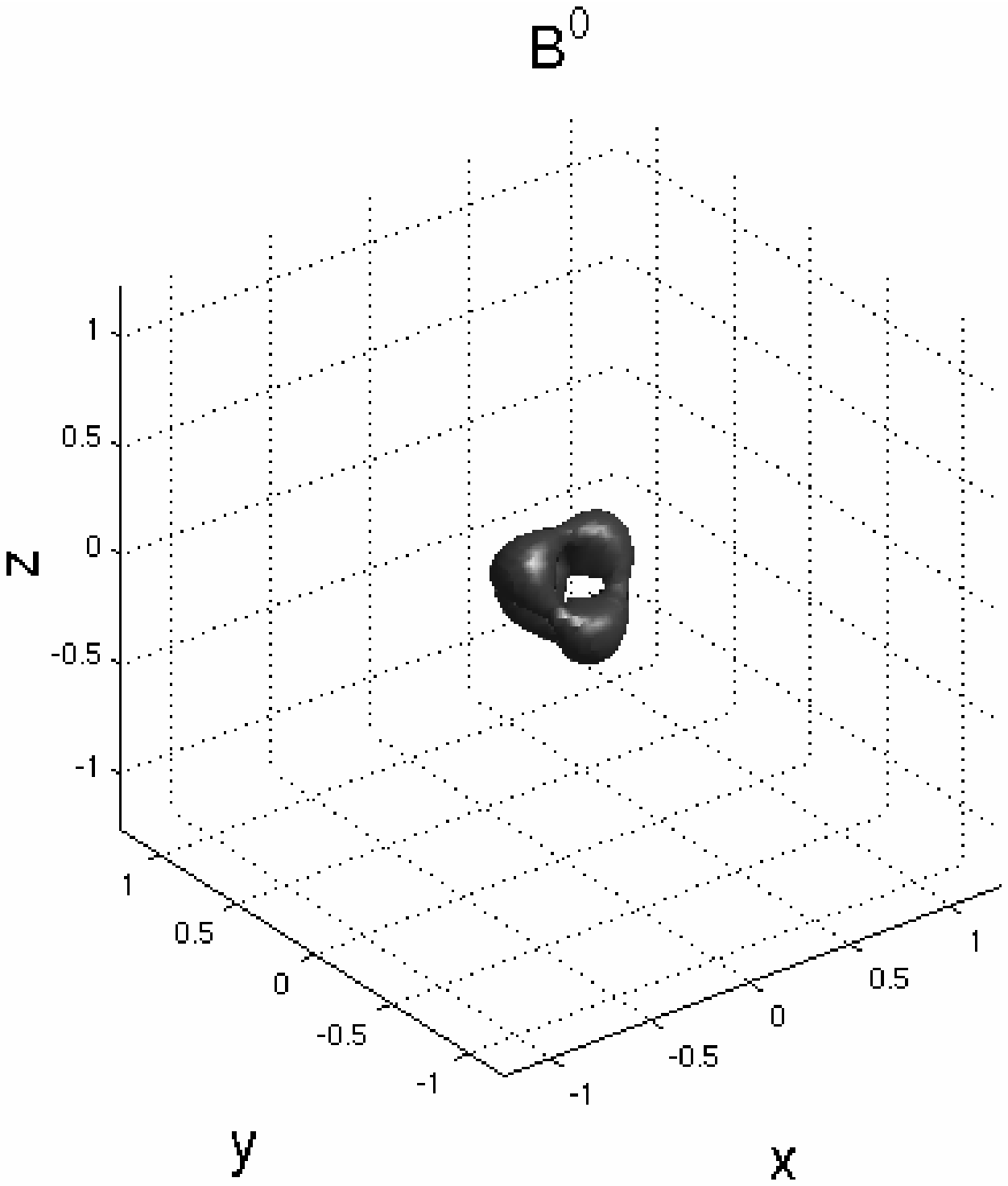} }
}\vspace{1.cm} }
\parbox{\textwidth}{
\centerline{
\mbox{\epsfysize=9.0cm \epsffile{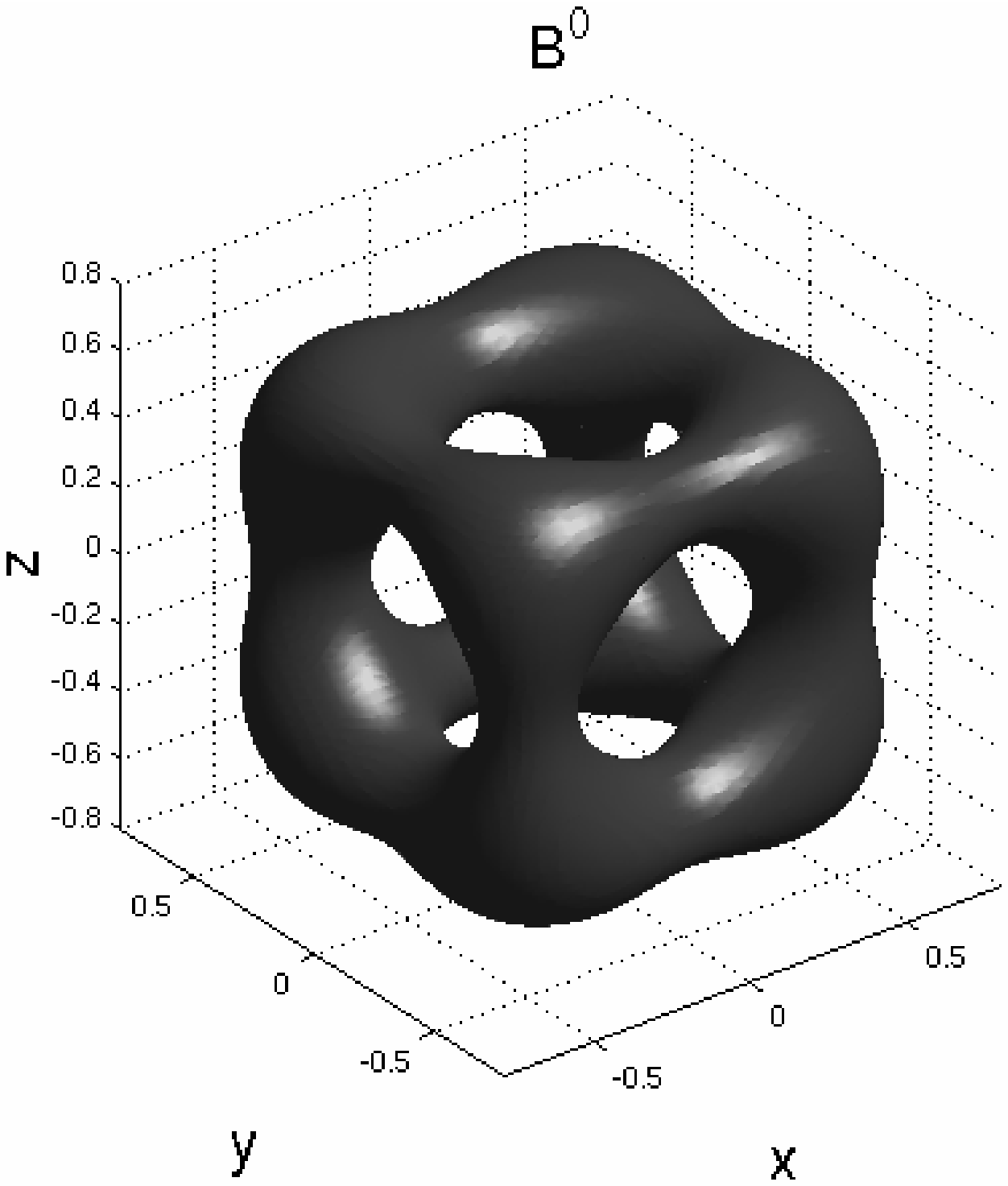} } \hspace{1cm}
\mbox{\epsfysize=9.0cm \epsffile{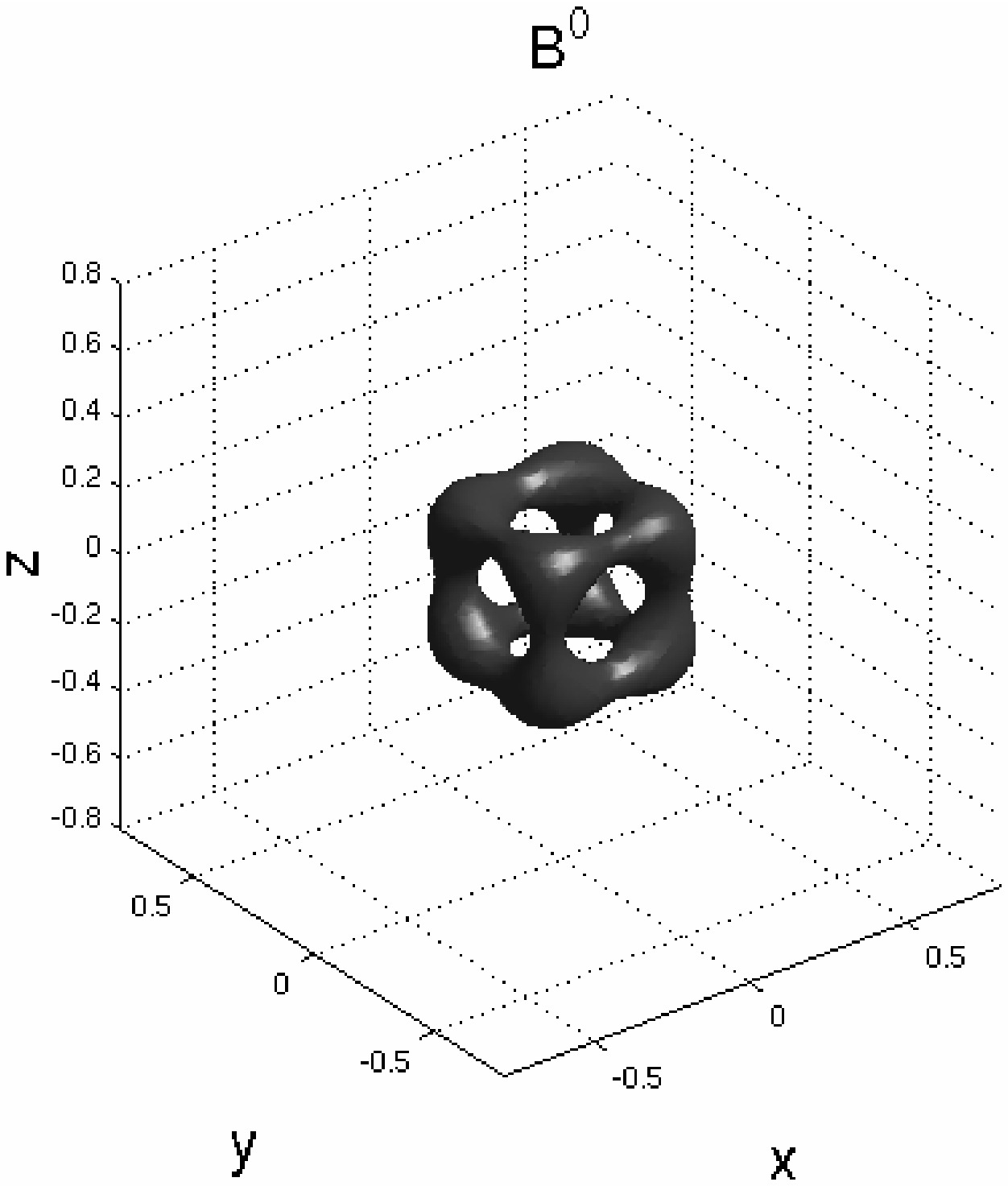} }
}\vspace{1.cm} }
\caption{\label{Fig3} 
Isosurfaceplot of the baryon density 
$B^0(x,y,z)=0.25 \times {\rm max}(B^0)$ for the $B=3$ Skyrmion
(upper row) and
$B^0(x,y,z)=0.5 \times {\rm max}(B^0)$ for the $B=4$ Skyrmion
(lower row)
in the ``$f-l$-approximation'' for $\alpha=0.02$
on the lower branch (left column) and the upper branch (right column).}
\vspace{0.5cm}
\end{figure}

Focussing now on the platonic Skyrmions, we exhibit in Figs.~\ref{Fig3}
surfaces of constant baryon density 
for tetrahedral $B=3$ and cubic $B=4$ Skyrmion solutions 
on the lower  branch (left) and upper branch (right). 
For a given rational map and coupling parameter $\alpha$
the Skyrmion on the upper branch is confined in a smaller volume
than the Skyrmion on the lower branch,
while the shape of the baryon density is
primarily determined by the rational map \cite{BTC},
analogous to the shape of the energy density \cite{KKM}.

\begin{figure}[t!]
\parbox{\textwidth}{
\centerline{
\mbox{\epsfysize=8.0cm \epsffile{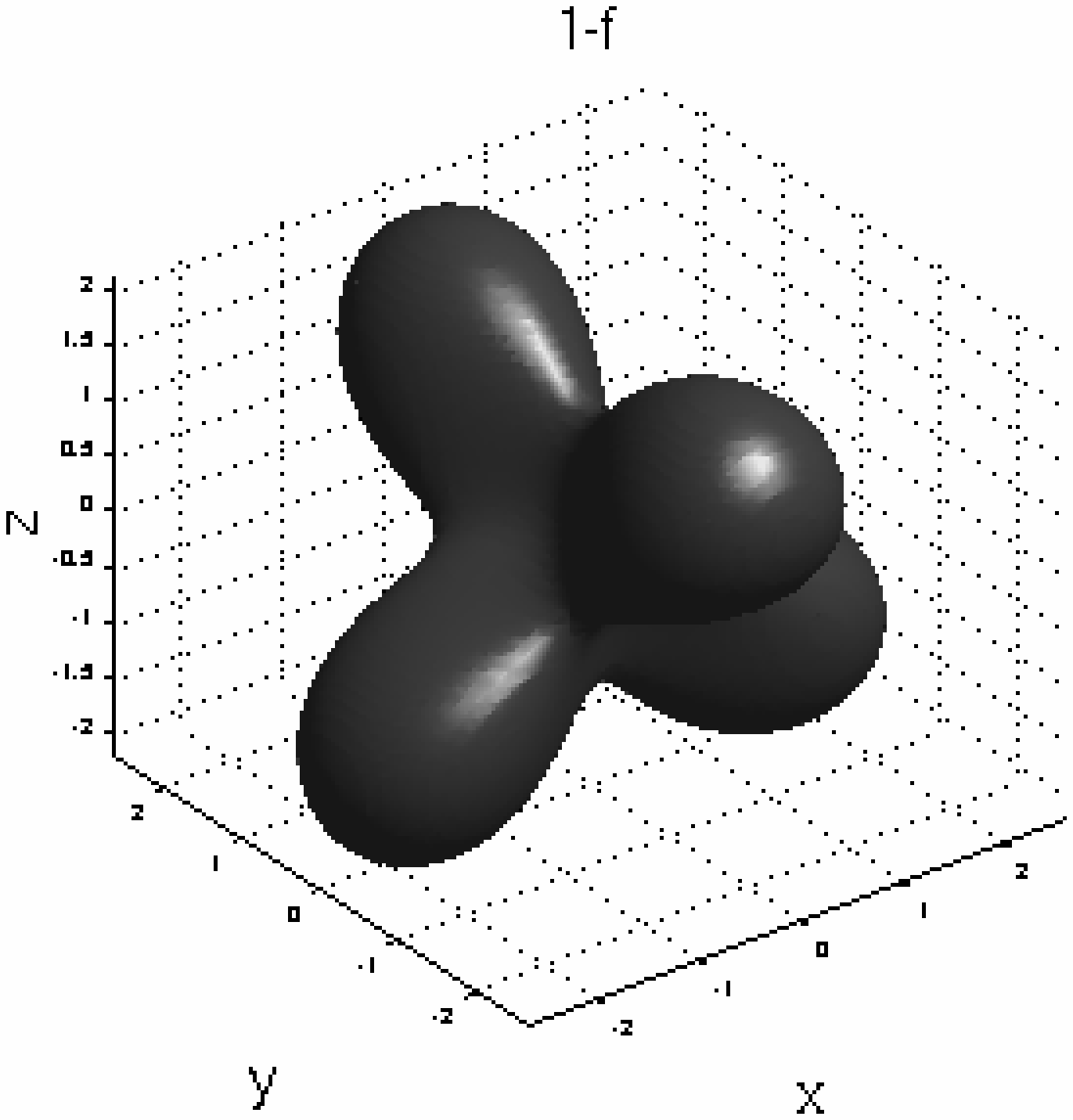} } \hspace{1cm}
\mbox{\epsfysize=8.0cm \epsffile{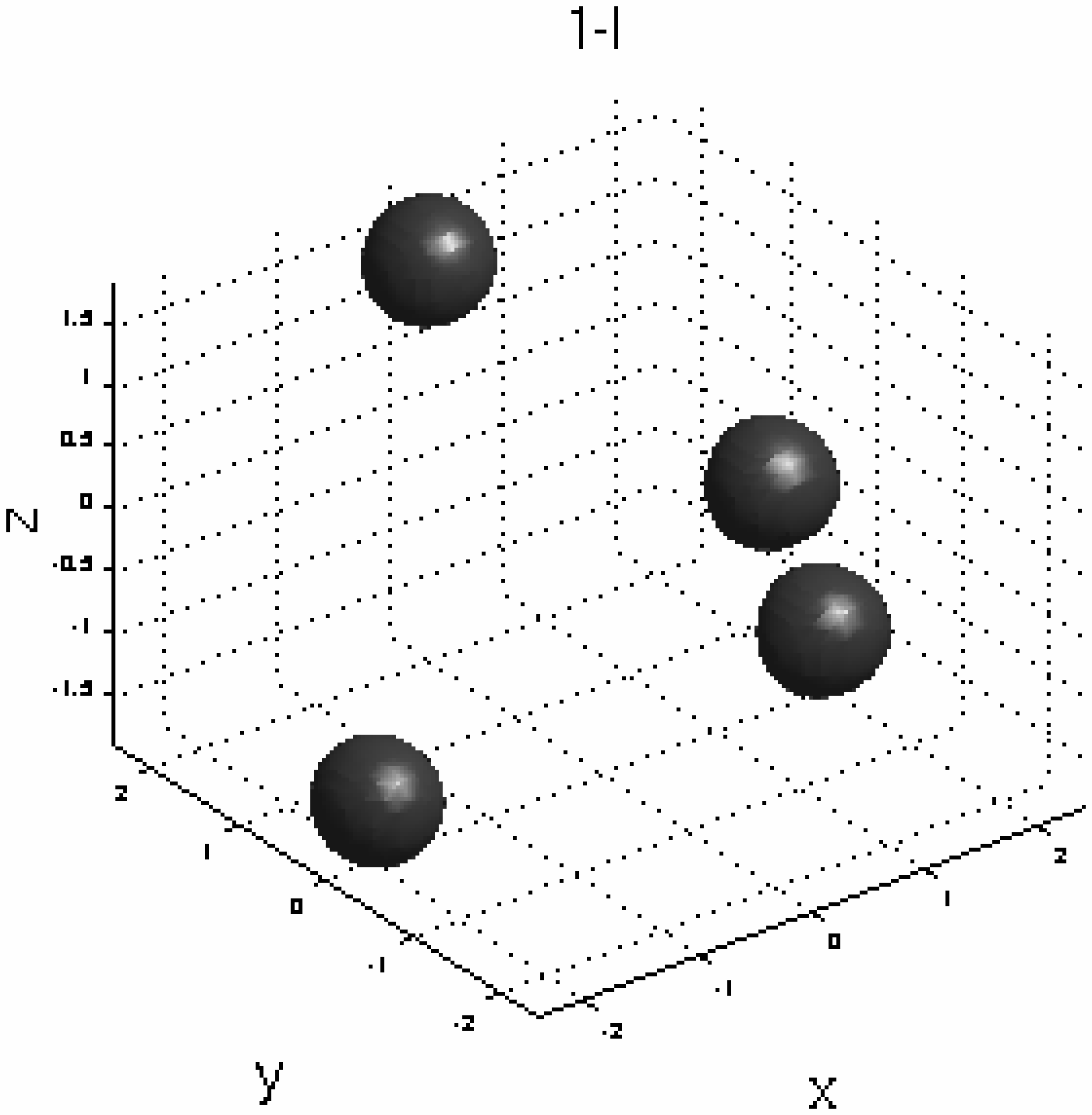} }
}\vspace{1.cm} }
\parbox{\textwidth}{
\centerline{
\mbox{\epsfysize=8.0cm \epsffile{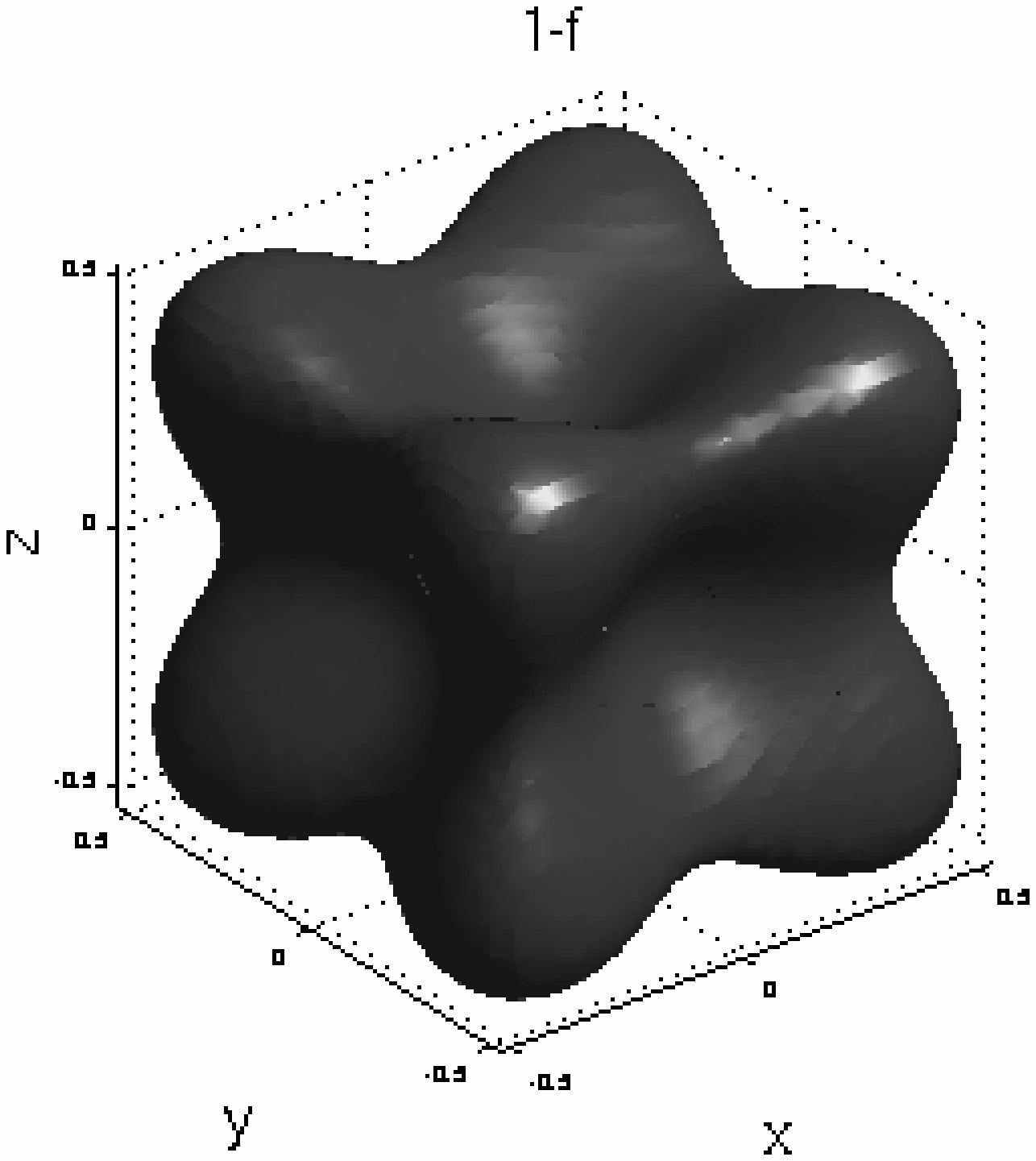} } \hspace{1cm}
\mbox{\epsfysize=8.0cm \epsffile{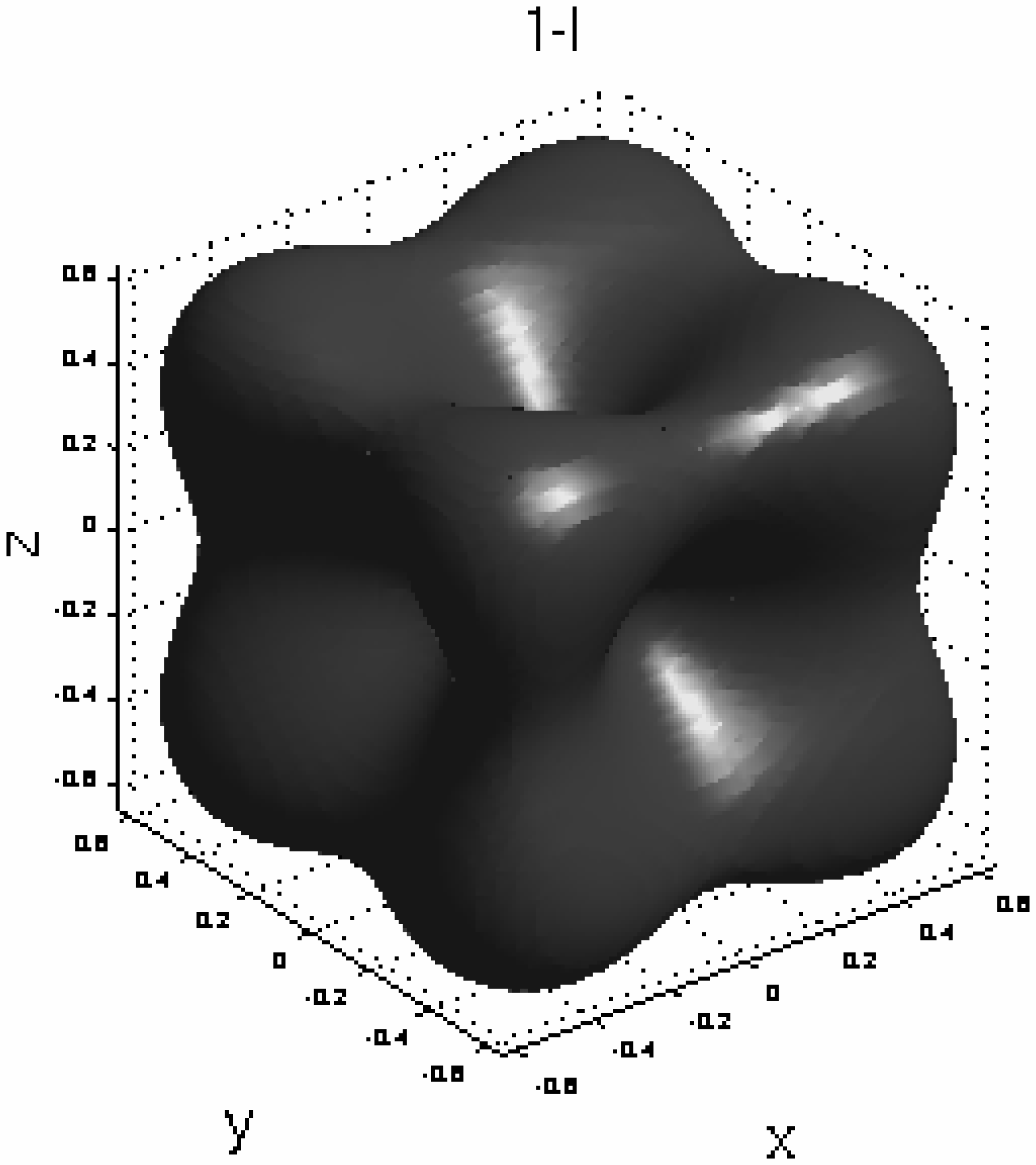} }
}\vspace{1.cm} }
\caption{\label{Fig4} 
Isosurfaceplots of 
the functions $1-f$ (left column) and $1-l$ (right column)  
for the $B=3$ Skyrmion (upper row) and the $B=4$ Skyrmion (lower row)
in the ``$f-l$-approximation'' for $\alpha=0.02$ on the lower branch.
} 
\vspace{0.5cm}
\end{figure}

In Figs.~\ref{Fig4} we demonstrate that the metric functions 
$f$ and $l$ of the platonic $B=3$ and $B=4$ Skyrmions
possess the same symmetry as the baryon density. 
In fact, 
when gravity is weak the function $1-f$ is proportional to the 
Newtonian gravitational potential. 

\section{Conclusions}

While aiming at the numerical construction
of exact platonic gravitating Skyrmions,
we have here obtained simpler approximate solutions,
based on the improved harmonic map ansatz
for the Skyrme field, thus avoiding 
the full complexity of the coupled
Einstein-Skyrme equations.
This ansatz for the Skyrme field involves a single function
instead of three functions.
Consequently, an appropriate
restriction of the ansatz for the metric is required,
involving either two functions (``$f-l$-approximation'')
or a single function (``dilaton-approximation'').

We have focussed on platonic gravitating Skyrmions
with tetrahedral ($B=3$) and cubic ($B=4$) symmetry.
For comparison, we have also constructed
gravitating axially symmetric Skyrmions ($B=2,3,4$).
The dependence on the coupling parameter $\alpha$
of the axial and platonic Skyrmions
is completely analogous to the $\alpha$-dependence of
spherical $B=1$ Skyrmions.
When $\alpha$ is increased from zero a lower branch of gravitating
Skyrmions emerges from the corresponding flat space Skyrmion solution.
This branch extends up to a maximal value $\alpha_{\rm max}$,
where it merges with an upper branch of solutions,
which extends back to $\alpha=0$.
Thus gravitating Skyrmions
exist only up to a maximal value of the coupling parameter,
$\alpha_{\rm max}$,
which depends on the specific rational map and on the approximation used.

The shape of the baryon density of platonic Skyrmions is
primarily determined by the rational map
and analogous to the shape of the energy density.
For a given rational map and coupling parameter $\alpha$
the Skyrmion on the upper branch is confined in a smaller volume
than the Skyrmion on the lower branch.
The metric functions of the platonic Skyrmions
possess the same symmetry as the baryon density. 

Comparing the approximations, applied in the construction 
of the gravitating tetrahedral and cubic Skyrmions,
we note that their mass is slightly higher in the ``$f-l$-approximation''
than in the ``dilaton-approximation''.
On the other hand, in flat space the exact Skyrmion \cite{BTC} 
has a slightly lower mass than the approximate Skyrmion \cite{IKZ2}.
Therefore, it is an open question whether the mass of the 
exact gravitating Skyrmion solutions
remains lower than the mass of the approximate solutions for
all values of the coupling parameter.
Construction of the exact platonic gravitating Skyrmion solutions, however,
remains currently still a numerical challenge.

{\bf Acknowledgement}: 

B.K.~gratefully acknowledges support by the DFG under contract
KU612/9-1.

\end{document}